\documentclass[reprint, amsmath,amssymb]{revtex4-2}

\usepackage{graphicx}% Include figure files
\usepackage{dcolumn}% Align table columns on decimal point
\usepackage{bm}
\usepackage{gensymb}
\usepackage[colorlinks=true,linkcolor=blue, citecolor=blue]{hyperref}
\graphicspath{{./fig_rb/}}
\usepackage{xcolor}

\newcommand{\alter}[1]{{\color{teal} #1}}

\begin{document}
	
	%\preprint{APS/123-QED}
	
	\title{Tuning photostriction in (PbTiO\ensuremath{_{3}})\ensuremath{_{n}}/(SrTiO\ensuremath{_{3}})\ensuremath{_{m}} superlattices via chemical composition: An \textit{ab-initio} study}% Force line breaks with \\

\author{Carmel Dansou\textsuperscript{1}, Charles Paillard\textsuperscript{1,2} ,Laurent Bellaiche\textsuperscript{1,3}}
\affiliation{\textsuperscript{1}Smart Ferroic Materials Center, Institute for Nanoscience \& Engineering and Department of Physics, University of Arkansas, Fayetteville, Arkansas 72701, USA}
\affiliation{\textsuperscript{2}Université Paris-Saclay, CentraleSup\'{e}lec, CNRS, Laboratoire SPMS, 91190 Gif-sur-Yvette, France.}
\affiliation{\textsuperscript{3}Department of Materials Science and Engineering, Tel Aviv University, Ramat Aviv, Tel Aviv 6997801, Israel.}

	\date{\today}
	
	\begin{abstract}
		Light-induced mechanical deformations in single-domain (PbTiO\ensuremath{_{3}})\ensuremath{_{n}}/(SrTiO\ensuremath{_{3}})\ensuremath{_{m}} superlattices were simulated using first-principle calculations. By varying the chemical fraction of PbTiO\ensuremath{_{3}}, we discover that these heterostructures' photostrictive behavior can be tuned quantitatively and qualitatively. Additionally, we present simple analytical models to explain the calculated deformations and predict a critical PbTiO\ensuremath{_{3}} fraction with no photostriction. In addition to the report in \cite{xxx}, our results present another way for tuning the photostrictive behavior of (PbTiO\ensuremath{_{3}})\ensuremath{_{n}}/(SrTiO\ensuremath{_{3}})\ensuremath{_{m}}  superlattices, which could be utilized for innovative optomechanical applications.
	\end{abstract}
	
	\maketitle
\section{Introduction}
In the last decade, ferroelectric/dielectric superlattices (FE/DE SLs) have gained significant interest because of their potential in next-generation multifunctional devices \cite{das20}. Among these FE/DE SLs, (PbTiO\ensuremath{_{3}})\ensuremath{_{n}}/(SrTiO\ensuremath{_{3}})\ensuremath{_{m}} have been widely studied as a model system~\cite{jiang1999,abid,ag1,ag2,bq,da1,da2,das,zub1,zub2,zubko20}. Many peculiar properties have been evidenced in such nanosystems, such as polar vortices and skyrmions or negative capacitance effects~\cite{abid,ag1,ag2,bq,da1,da2,das,zub1,zub2,zubko20}. Recently, significant photostriction - the light-induced mechanical deformation in a material - has been evidenced in these superlattices \cite{ahn2017,lee2021,dar}, adding yet another dimension to the rich number of their functionalities. Yet, there is little microscopic understanding of the origin of this effect in FE/DE SLs. There exists thus an urgent need to develop models and understanding of the photo-induced deformation in FE/DE SLs. 

In a parallel report \cite{xxx}, we show how the electrostatic coupling strength between the layers of a PTO/STO SL with fixed chemical composition can strongly change the photostrictive response from a compression to an extension by varying the overall SLs period from 2 to 10 perovskite unit cells. In the present report, we demonstrate how the chemical composition, at fixed SLs period, can also strongly impact the amplitude and sign of the photostriction. 
Specifically, we present a first-principle study of the photostrictive behavior of PbTiO\ensuremath{_{3}})\ensuremath{_{n}}/(SrTiO\ensuremath{_{3}})\ensuremath{_{m}} SLs ($m+n=10$). Here, we generalize and extend the discussion of the results presented in \cite{xxx}, changing the chemical period of the SLs. Our aim is to provide, in addition to the results in \cite{xxx}, a general theoretical understanding into the mechanisms and ways to tune photostriction in these SLs with any chemical composition. In the present report, we aim at uncovering the dependence of photostriction in PTO/STO SLs on the PTO volume fraction. Our report is organized as follows. Construction of the simulated PTO/STO SLs and the computational methods are detailed in Sec \ref{s1}. In Sec \ref{s2}, we briefly summarize the results from \cite{xxx} on $(n|n)$ SLs. In Sec \ref{s3}, we present and analyze the photostrictive behavior of the $(n|m)$ SLs as function of PTO ratio. Section \ref{s4} is concerned with phenomenological models to quantify light-induced strains in the SLs. And we close this report by a summary of our findings and future perspectives.
We thus hope that the photostrictive tunability of PTO/STO SLs explored in this article may be of significant interest to the application of FE/DE SLs for optomechanics application and may open new avenues for fundamental explorations.

\section{Systems and Methods}
\label{s1}	
\subsection{Computational details}
Calculations in this report are based on density functional theory (DFT) using the projector augmented wave method \cite{paw0} as implemented in the Abinit package \cite{abinit1,abinit2,abinit3}. The Perdew-Burke-Ernzerhof generalized gradient approximation revised for solids is used for the exchange-correlation functional \cite{perdew}. The plane-wave energy cutoff is set to $20$~Hartree. Integration over the Brillouin zone is done using a Monkhorst-Pack \cite{monk} scheme equivalent to $6\times 6 \times 6$ k-point mesh in a five-atom perovskite cell. To simulate the $c$ phase, the Ti atoms are displaced up along the z axis from their centrosymmetric positions, and the out-of-plane cell dimension, atomic positions and angles are relaxed until forces on the supercell are less than $10^{-8} \text{ Hartree.Bohr}^{-1}$ and the force on ions less than $2\times 10^{-6}\text{ Hartree.Bohr}^{-1}$. At the end of the relaxation, the obtained structures have the {\it P4mm} space group symmetry with polarization along the $+\hat{z}$ direction. Starting from the optimized structure in the dark, photoexcitation is simulated using the two quasi-Fermi levels method to mimic thermalized electrons and holes recently implemented in the Abinit \cite{paillard2019} code by constraining a fixed number $n_{ph}$ of electrons/holes to the conduction/valence bands. Under the constraint of  $n_{ph}$ electrons/holes in the conduction/valence bands, the structures were relaxed again with the same convergence criteria \alter{as} in the dark ($n_{ph}=0$).

\subsection{Systems}
\begin{figure*}[ht!]
	\centering
	\includegraphics[width=1\linewidth]{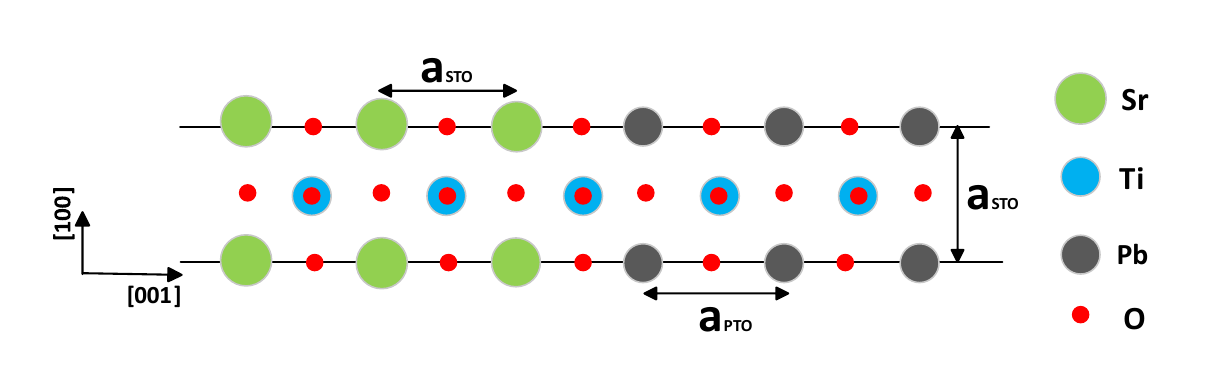}
	\caption{\label{f1} Schematic of the simulated PTO/STO superlattices. The atoms are shown in their perfect centrosymmetric positions.}
\end{figure*}

To simulate the PTO/STO SLs, we adopt a supercell approach used in many previous studies on FE/DE SLs \cite{ag1,ag2,bq,brist,gu2010}. Stoichiometric SLs are built by stacking alternatively $n$ unit cells of PTO on top of $m$ unit cells of STO with $n+m =10$ along the [001] direction, corresponding to a total of $50$ atoms in our simulation box. To construct the SLs, we first optimized the lattice parameter of both PTO and STO. For cubic STO, the optimized lattice value is $3.897$~\AA~in good agreement with existing experimental value $3.905$~\AA~\cite{stos1,stos2,stos3} and previously computed values \cite{stot1}. For cubic PTO, we found $a=3.9248$ \AA~which also is in good agreement with experimental \cite{ptos1,ptos2,ptos3} and previously reported theoretical values \cite{ptot1,ptot2,pail2}. For tetragonal PTO, we found $a=3.885$ \AA~ and $c/a=1.0712$ well comparable to experimental values $3.902$ \AA~($c/a=1.0650$) \cite{ptos1,ptos2,ptos3} and previous computational studies $3.888$ \AA~and $c/a =1.0739$ \cite{ptot1,ptot2}. Before structural relaxation, atoms are placed in their ideal cubic positions, and the out-of-plane lattice constant of the supercell is set to ($ma_{\text{STO}}+na_{\text{PTO}})$, with $a_{\text{STO}}$ and $a_{\text{PTO}}$ being the theoretical optimized values of cubic PTO and STO, respectively. A schematic of the SLs is shown in Fig \ref{f1}. Only SLs with polarization along the [001] direction commonly denoted the $c$ phase in literature are considered in this study \cite{pertsev}. Structural, electronic and physical properties of the c phase of PTO/STO SLs are well documented \cite{ag1,gu2010,da1,da2,zub1,zubko20,zhen2007,eglitis2016}. Additionally, we build reference paraelectric phases for the calculation of the layer-by-layer polarization. We aim at simulating PTO/STO SLs grown on an STO substrate as commonly done in experiment. We therefore fix the in-plane lattice constant of the SLs to the theoretical computed value for STO. We did not consider oxygen octahedral tilts modes in our simulation as they would require doubling the cell along the in-plane directions making our simulations computationally involved especially during photoexcitation. Also only single domain SLs are considered in this study.
\section{Results from the ($n|n$) PTO/STO Superlattices}
\label{s2}
To give context to the results presented in this report, we first summarize the findings presented in the companion paper \cite{xxx}. In \cite{xxx}, we investigated the light-induced deformation of single domain PTO/STO SLs with equal PTO and STO chemical fraction denoted $(n|n)$ SLs, $n$ varying from $1$ to $5$. We show that, for very thin SLs ($n=1-2$), free carriers are mostly delocalized and screen existing dipoles, which induces contractions in the out-of-plane lattice of the SLs. On the other hand, when the film is thick enough ($n=3-5$), the free charges are strongly localized in the interface regions. They produce a field that opposes existing fields (polarizing field in STO and depolarizing field in PTO) in the SLs and thereby induces a contraction in the STO layer and oppositely an expansion in the PTO layer.  In the SLs, we found the scaling $P_{z} \propto \eta_{T}^{\alpha} = \left[ (c-c_{p})/a \right]^{\alpha}$ relation ($c$ and $c_{p}$ are the out-of-plane lattice constants of the polar SLs and the corresponding paraelectric SLs, and $a$ the in-plane lattice parameter of the SLs), consistent with earlier reports \cite{lee2021,da2}. This relation enables us to relate the light-induced strains to the relative light-induced change in the polarization as follows:
\begin{equation}\label{eq1}
	\eta_{33} = \dfrac{\eta_{d}}{\alpha} \left( \dfrac{\Delta P_{z}}{P_{z}}\right);
\end{equation}
with $\eta_{d} = (c-c_{p})/c $. As shown in \cite{xxx}, the estimated light induced strain using Eq. \ref{eq1} agrees well with the DFT data.
\section{Photostriction as function of PTO volume fraction}
\label{s3}
\begin{figure}[ht!]
	\centering
	\includegraphics[width=1\linewidth]{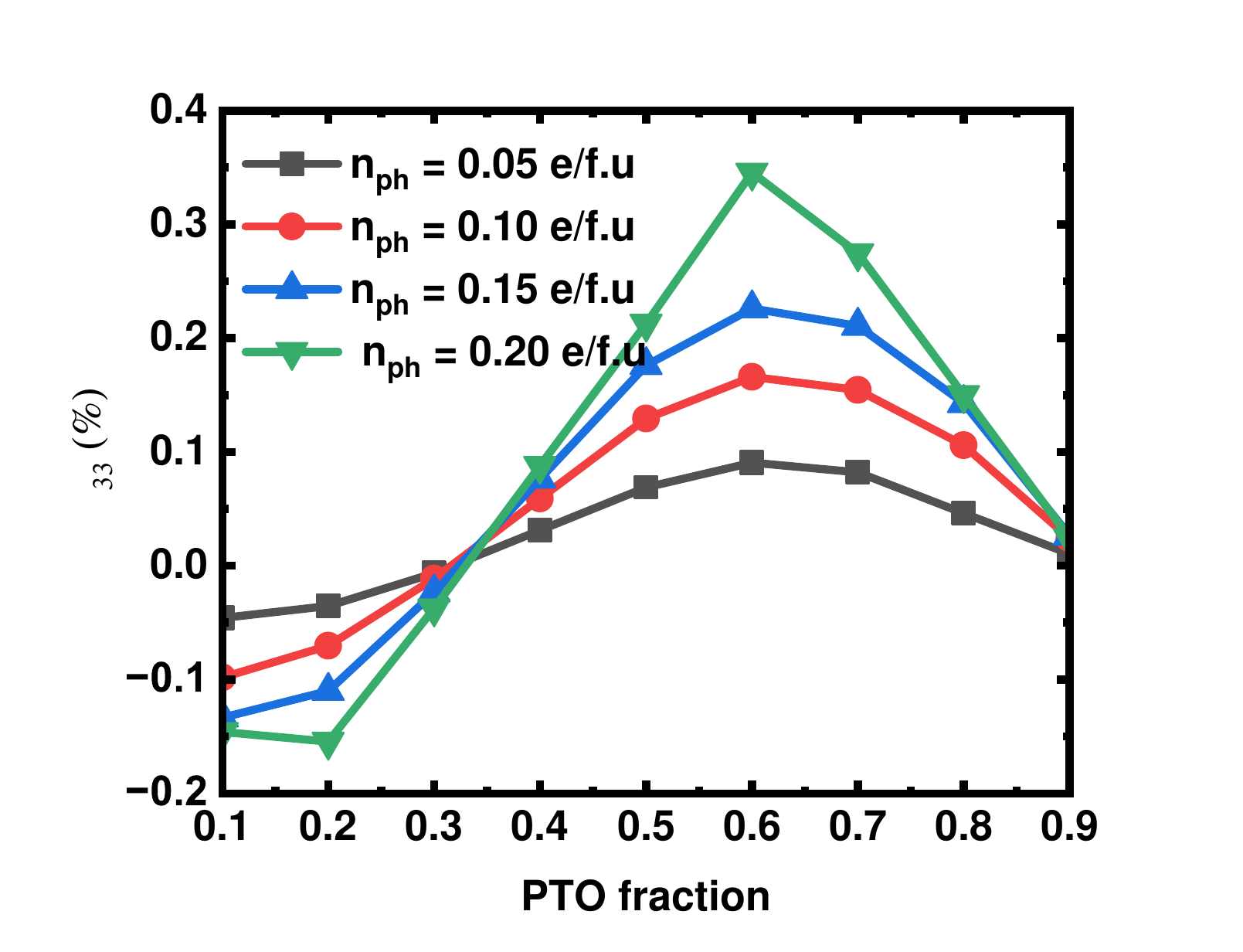}
	\caption{\label{f2} Light-induced strains as function of PTO fraction in the $(n|m)$ SLs.}
\end{figure}

We start by discussing the photostrictive behavior of the $(n|m)$ SLs as a function of the PTO fraction. As shown in Fig. \ref{f2}, the photostrictive behavior of the $(n|m)$ SLs  is dependent on the PTO fraction. 
Like for the $(n|n)$ SLs \cite{xxx}, two behaviors can be distinguished when we look at the sign of the light-induced deformation in the out-of plane lattice of the SLs that we denote as $\eta_{33} = (c-c_{0})/c_{0}$, where $c$ is the out-of-plane lattice constant of the SLs under illumination and $c_{0}$ its out-of-plane lattice constant in the absence of illumination.
\begin{figure}[ht!]
	\centering
	\includegraphics[width=1\linewidth]{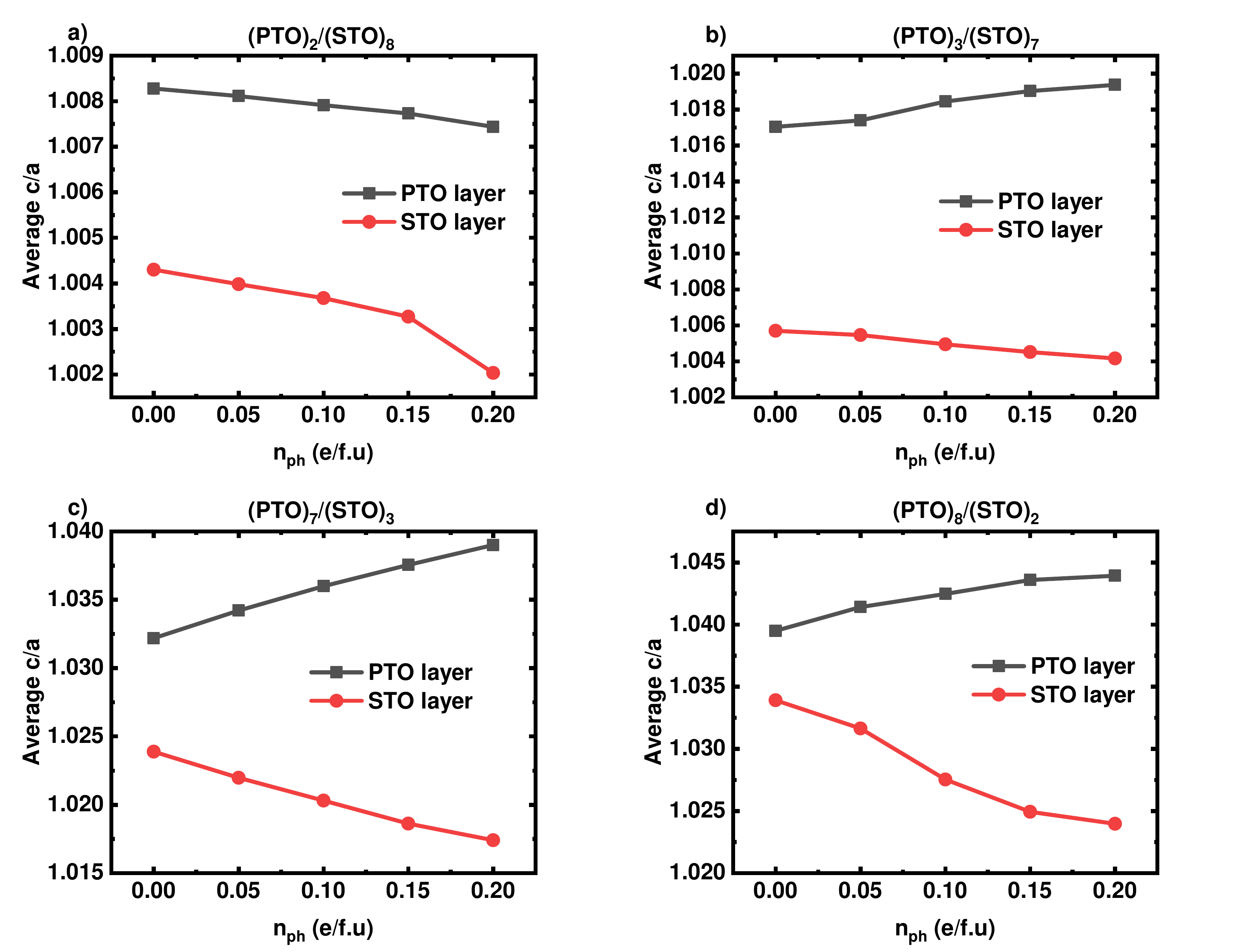}
	\caption{\label{f3} a)  Average c/a for PTO and STO layer in the $(2|8)$ SLs, b)  Average c/a for PTO and STO layer in the $(3|7)$ SLs, c)  Average c/a for PTO and STO layer in the $(7|3)$ SLs, d)  Average c/a for PTO and STO layer in the $(8|2)$ SLs  as function of $n_{ph} = 0.0-0.2$ e/f.u.}
\end{figure}
As shown in Fig. \ref{f2}, for PTO fractions less than a critical fraction $x_{c}\approx 0.34$ (estimated from the graphs on Fig. \ref{f2}; this value is also independent of $n_{ph}$), the SLs shows negative photostriction ($\eta_{33} < 0$) while for fraction between $x_{c}$ and $0.9$, the SLs shows positive photostriction ($\eta_{33} > 0$). Note also that beyond $x_c$, the light-induced deformation in the SLs, for each considered $n_{ph}$, as function of PTO fraction, reaches a maximal value at $x=0.6$ and subsequently decreases back to negative values in pure bulk PTO, consistent with previous calculations~\cite{pail1,pail2}. The layer-by-layer decomposition of the photo-induced average tetragonality ($c/a$) shows that, for PTO fraction $x \leq 0.2$, optical excitation penalizes distortions in both PTO and STO layers (see Fig \ref{f3}a where we show data for the $(2|8)$ SLs), similar to $(1|1)$ and $(2|2)$  SLs \cite{xxx}. On the other hand, for PTO fraction $x \geq 0.3$, the free carriers decrease distortion in STO layer while they enhance distortion in the PTO layer (see Figs \ref{f3}b, c\& d for selected SLs, see \cite{sprb} for the other SLs), similar to our observation in the $(n|n)$ SLs with $n\geq 3$  

We estimate the polarization in the SLs in a layer-by-layer fashion as done in the companion paper \cite{xxx}.
\begin{figure}[ht!]
	\centering
	\includegraphics[width=1\linewidth]{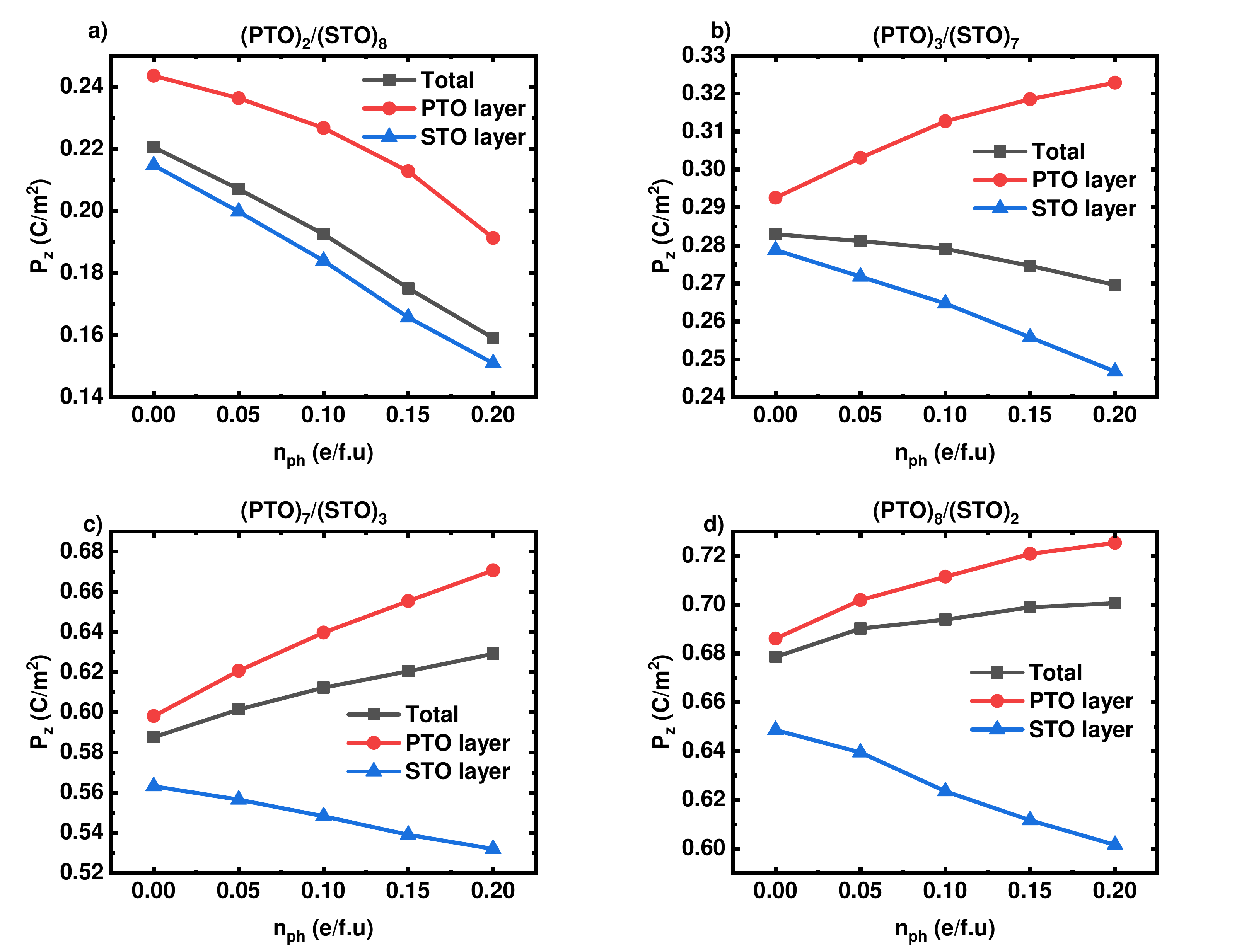}
	\caption{\label{f4} Polarizations as function of number of excited carriers for the a) $(2|8)$ SLs, b) $(3|7)$ SLs, c) $(7|3)$ SLs, d) $(8|2)$ SLs.}
\end{figure}
\begin{figure}[ht!]
	\centering
	\includegraphics[width=1\linewidth]{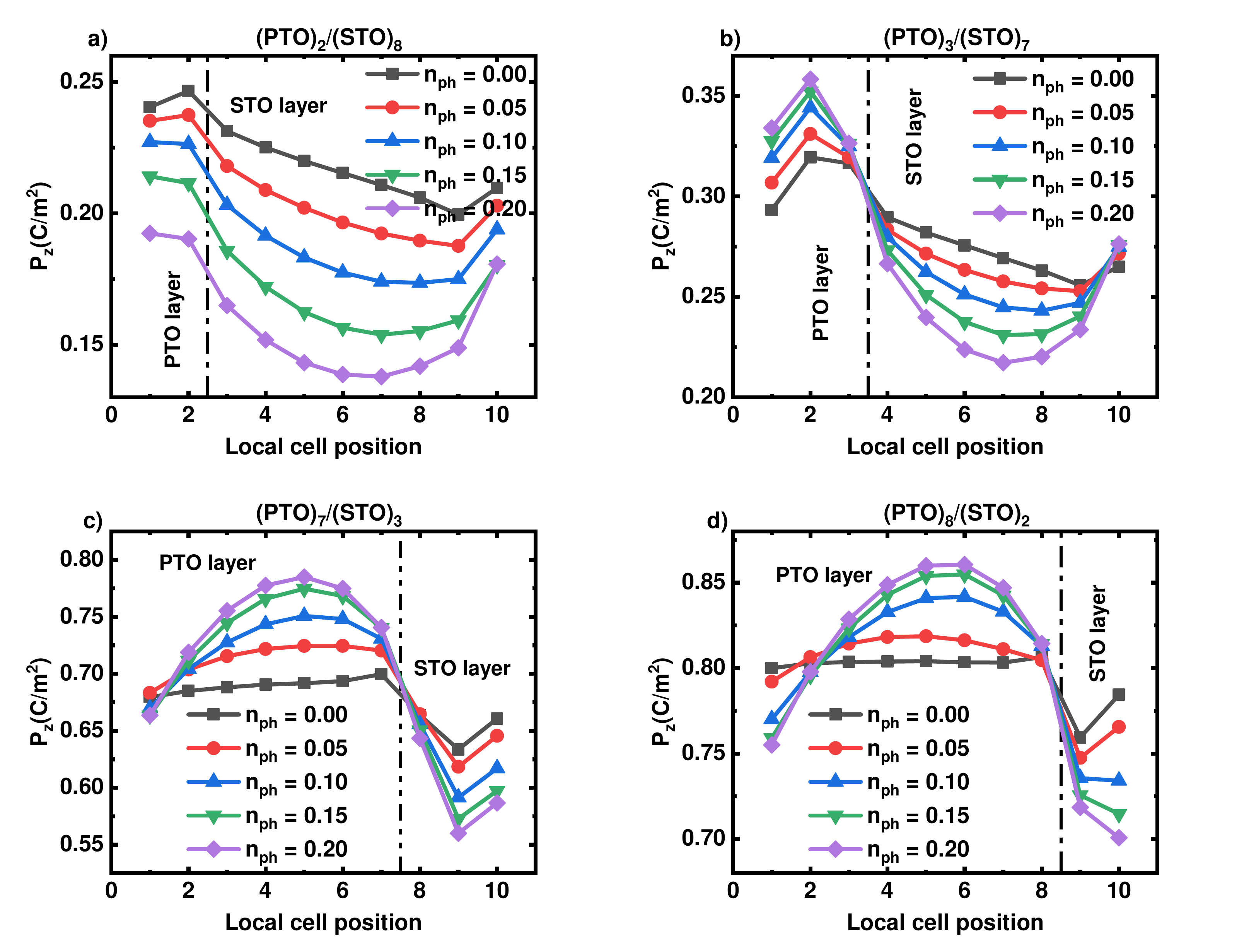}
	\caption{\label{f5} Layer-by-layer decomposition of the polarizations as function of number of excited carriers for the a) $(2|8)$ SLs, b) $(3|7)$ SLs, c) $(7|3)$ SLs, d) $(8|2)$ SLs. are to separate local cells corresponding to PTO and STO layer in each SLs.}
\end{figure}
On Fig \ref{f4}, we show the estimated total polarization as well as polarization in each layer for selected SLs (See  \cite{sprb} for the rest of the SLs) as function of $n_{ph}$. The decomposed polarization per local five atoms unit cell of the SLs are also shown in Fig \ref{f5} (we only show for selected SLs, see the rest in the supplemental materials). For PTO fraction $x \leq 0.2 $, optical excitation reduces polarization in both PTO and STO layer and thus the reducing of the total polarization in the SLs, see Figs  \ref{f4}a and Figs \ref{f5}a. When the fraction of PTO is increased above $0.2$ ($x \geq 0.3$), polarization in STO is decreased by optical excitation whereas polarization in PTO is enhanced, see Figs \ref{f4} and Figs \ref{f5} (b,c,d). The total polarization of the SLs decreases for $x \leq 0.3$ and increases for $x\geq 0.4$ PTO fraction. These light-induced change in the polarization in the SLs are in line with the observations from the lattice distortion in the SLs discussed in the precedent section.
The overall behaviors (polarization and lattice distortion) of these SLs is thus the resultants of two actions (the total strain in the SLs is the resultant of the light-induced strain in PTO layer and STO layer as shown in the piezoelectric model in section \ref{s4}B). This seems to be a general feature of FE/DE SLs, since a recent experiment on $(2|4) $ BaTiO\ensuremath{_{3}}/CaTiO\ensuremath{_{3}} SLs showed that optical excitation induced compression in CaTiO\ensuremath{_{3}} and oppositely an expansion in BaTiO\ensuremath{_{3}} \cite{sri202}. 
\section{Models and discussions}
\label{s4}
\subsection{First Model: $P_{z} \propto \eta_{T}^{\alpha} $}
To describe the light-induced mechanical response within a phenomenological method, we applied the $P_z \propto \eta_{T}^{\alpha}$ model, described in our companion paper \cite{xxx}, to each of the SLs. In this phenomenological model, the light-induced strain is related to the relative change in the polarization via Eq. \ref{eq1}. For each of the SLs, we obtained the value of $\alpha$ by a linear fit of $\ln(P_{z}(n_{ph}))~ vs ~\ln(\eta_{T} (n_{ph}))$ as described in \cite{xxx} (see \cite{sprb} for the values of $\alpha$).  On figure \ref{f7}, we show the results (for selected SLs, the results for other SLs can be found in the  Supplemental Material \cite{sprb}) of our fit along with the estimation of the strain using $\alpha = 0.5$ and $\alpha = 1.1$ for comparaison.
\begin{figure}[ht!]
	\centering
	\includegraphics[width=1\linewidth]{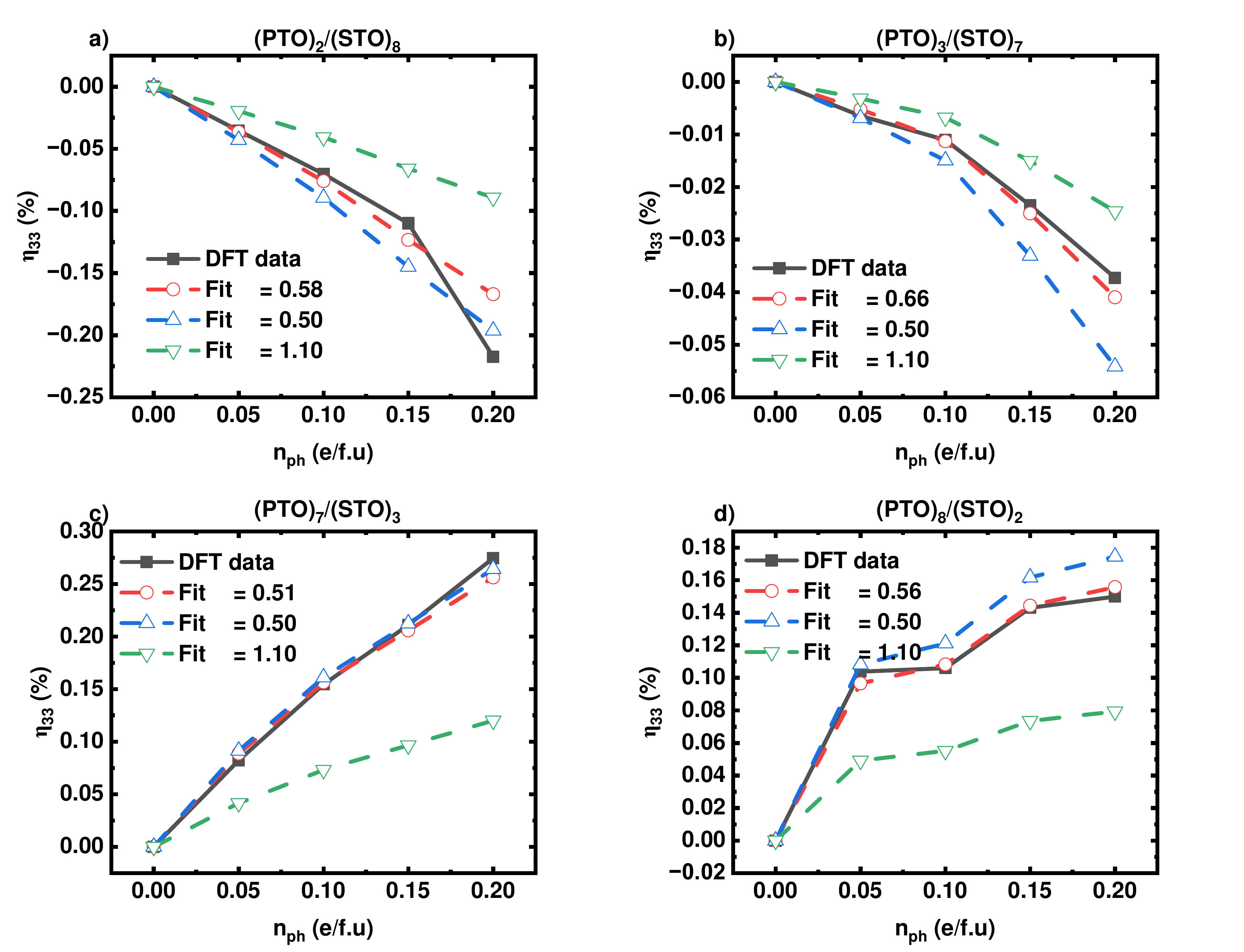}
	\caption{\label{f7} Different fits of the DFT data ($\eta_{33}~ vs~ n_{ph}$) using Eq \ref{eq1} for a) $(2|8)$ SLs, b) $(3|7)$ SLs, c) $(7|3)$ SLs.}
\end{figure}
The model with estimated $\alpha$ from the scaling relation fits better the DFT data for the corresponding SLs. As discussed in \cite{xxx} and by Lee {\it et al.} \cite{lee2021}, we are unable to attach a physical meaning to $\alpha$ and its changes as function of photoexcited charges and the type of SLs. We leave this as an interesting future exploration.
\subsection{Converse piezoelectric effect}	
\begin{figure}[ht!]
	\centering
	\includegraphics[width=1\linewidth]{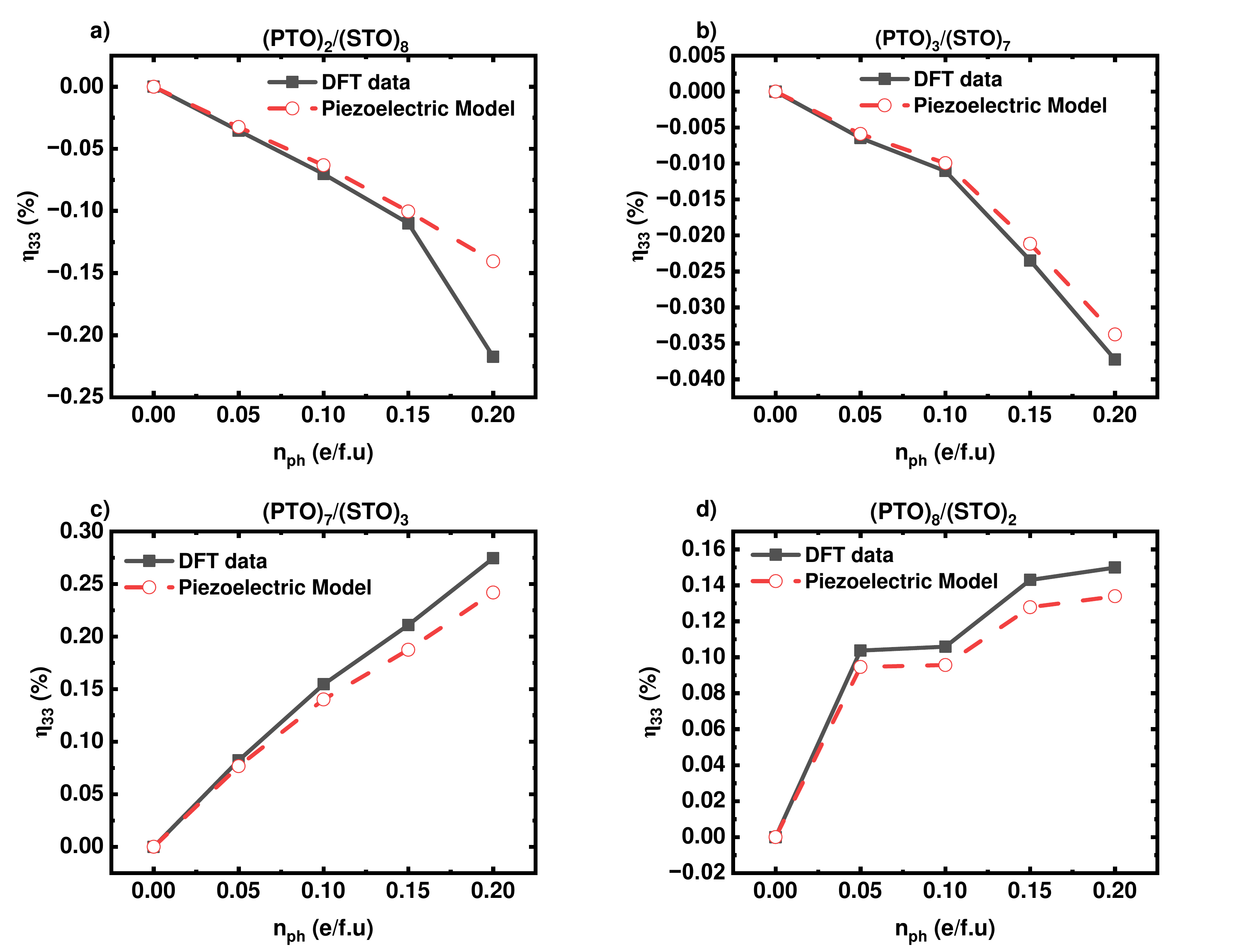}
	\caption{\label{f8} Different fits of the DFT data ($\eta_{33}~ vs~ n_{ph}$) using Eq \ref{eq3} for a) $(2|8)$ SLs, b) $(3|7)$ SLs, c) $(7|3)$ SLs, d) $(8|2)$ SLs.}
\end{figure}

Here, we aim at finding another simple model to describe and reproduce our DFT data. The total light-induced strain can be approximated as the weighted average of the deformation in STO and PTO layers in the SLs (see  Supplemental Material \cite{sprb} for details and discussion of the derivation):
\begin{equation}\label{eq2}
	\eta_{33} \approx x \delta \eta_{33}^{p} +\left(1-x\right)\delta\eta_{33}^{s} 
\end{equation}
where $x$ is the volume fraction of PTO in the SLs. $\delta \eta_{33}^{p}$ is the light-induced strain in the PTO layer and $\delta\eta_{33}^{s}$ the light-induced strain in the STO layer. The volume fraction $x$ changes as function of $n_{ph}$. Introducing the changes in the polarization ($\delta P_{1}= P_{1}(n_{ph} \neq 0) - P_{1}(n_{ph} = 0)$) for PTO and  ($\delta P_{2}= P_{2}(n_{ph} \neq 0) - P_{2}(n_{ph} = 0)$) for STO), and writing the following converse piezoelectric relation between strain and change in polarization \cite{pail2}:

\begin{equation}
	\delta \eta_{33}^{p} =\epsilon_{0}^{-1}  \dfrac{d_{33}^{1}}{\chi_{33}^{1}}  \delta P_{1}; ~~ \delta\eta_{33}^{s}=\epsilon_{0}^{-1} \dfrac{d_{33}^{2}}{\chi_{33}^{2}} \delta P_{2},
\end{equation}
we have: 
\begin{equation}\label{eq3}
	\eta_{33} \approx \epsilon_{0}^{-1} \left[ x \left(\dfrac{d_{33}^{1}}{\chi_{33}^{1}}  \delta P_{1} - \dfrac{d_{33}^{2}}{\chi_{33}^{2}} \delta P_{2} \right) + \dfrac{d_{33}^{2}}{\chi_{33}^{2}} \delta P_{2} \right];
\end{equation}
where $(d_{33}^{1},\chi_{33}^{1}) $ and $(d_{33}^{2},\chi_{33}^{2})$ are the coupled piezoelectric constants, dielectric susceptibility of the PTO layer and the STO layer respectively. Note that although STO is cubic in its bulk structure, it becomes tetragonal, with a finite internal field, in the SLs due to its proximity with the PTO layer and thus has a nonvanishing piezoelectric constant $d_{33}$. The quantities $d_{33}^{i}$, and $\chi_{33}^{i}$ change with illumination and their values can be computed, in principle, within density functional perturbation theory (DFPT). However, the current implementation of DFPT in Abinit does not account for the presence of free charge in the conduction band. As such, we define the ratio $r^{i} = d_{33}^{i}/\chi_{33}^{i}$ and estimate $r^{i}$ from the DFT data. The estimation of $r^{i}$ are detailed in the Supplemental Materials \cite{sprb}. Eq. \ref{eq3} can therefore be rewritten as:
\begin{equation}\label{eq4}
	\eta_{33} \approx \epsilon_{0}^{-1} \left[ x \left( r^{1}  \delta P_{1} - r^{2} \delta P_{2} \right) + \delta P_{2} r^{2} \right]
\end{equation}
As observed in the DFT data, this model predicts a critical PTO fraction for a critical SLs that is at the boundary of negative and positive photostrictive SLs:
\begin{equation}
	x_{c}  =  \dfrac{1}{1-\dfrac{\delta P_{1}}{\delta P_{2}} \dfrac{r^{1}}{r^{2}}}
\end{equation}
Interestingly, the DFT data suggests that this critical volume fraction is independent of $n_{ph}$. On Fig \ref{f8}, we show the results from this model and the DFT data for $n_{ph}=0.00-0.20$ e/f.u for few selected SLs (the rest of the SLs is shown in the Supplemental Material \cite{sprb}). For such a simple model, it agrees rather well and captures the underlying physics. The model, however, fails to quantitatively capture the light induced strains at higher $n_{ph}$ for some of the SLs. This is because at high $n_{ph}$, it can be noticed that the photostrictive behavior of some of the SLs becomes very nonlinear and a rigorous calculation of the $r^{i}$ ratio in Eq. \ref{eq4} with DFPT might improve the model at large $n_{ph}$. 
We also would like to relate this study to the two only existing experimental studies we are aware of on light-induced strain in PTO/STO SLs. In Refs. \cite{ahn2017,lee2021},  a film of 8 units of PTO and 3 units of STO was used. This corresponds to roughly a $0.72$ PTO fraction. They all reported an overall positive photostrictive behavior, qualitatively consistent with the observation in our study. 

\section{Conclusion and perspectives}
In summary, using DFT within a constrained occupation scheme, we investigate photostriction in single domain PTO/STO SLs with varying chemical composition. We find that, for PTO fraction $x\leq 0.2$, both layers of the SLs contract upon optical excitation, leading an overall contraction of the SLs. On the other hand, for $x\geq 0.3$ , optical excitation induces a contraction in the STO layers and an expansion in the PTO layers. Depending on the fraction of PTO in the SLs, an overall contraction or expansion is observed. This study, in conjunction with the report in \cite{xxx}, show that the photostrictive behavior of these SLs can be tuned not only through the SLs composition but also its thickness. The possibility of tuning the light-induced strain by simply changing the SLs composition or thickness is appealing as it presents potential for novel applications using light as modulator.

In the present study, as well as in the one in  \cite{xxx}, we restrict ourselves, because of the computational cost of simulating optical excitation in large system, to single domain PTO/STO SLs with no octahedral tilts (oxygen octahedral tiltings only appear in STO below 105K \cite{scott1974soft} and does not exist in pure PTO. Additionally existing experimental studies\cite{ahn2017,lee2021,dar} were done at much higher temperature and did not consider any octahedral tiltings, thus neglecting their effect is a realistic consideration), consisting of a total of $10$ five-atom perovskite cells. Consequently, this choice limits the predictive power of our studies and a quantitative comparison with reported experimental values in Refs.\cite{ahn2017,lee2021} can be misleading. However, note that our results are qualitatively in line with these reports. We believe that, through these combined investigations, we have provided first-principle insights into the photostrictive behavior of PTO/STO SLs. Additionally, our studies can be used by experimentalists as a starting point to optimize the photostrictive response of PTO/STO SLs. A straightforward future development is to extend these studies to thicker SLs (i.e., SLs with more than $10$ five-atoms perovskite cells) and investigate how the dependence on PTO chemical fraction of the photostrictive behavior (shown on Fig \ref{f2}) scales with the SLs thickness. Another approach would be to extend this study to SLs with domain structures, octahedral tilts or super-orders \cite{xu2024,gu2024}. Expanding on the ideas presented in our reports in these directions offers an interesting avenue for future explorations. However, pursuing these ideas may require treating the ``constrained occupation scheme" in an effective way through large scale methods such as effective Hamiltonians or second-principles approaches, as simulating photoexcitation in these large systems using DFT becomes computationally cumbersome.

\begin{acknowledgements}
	The work is supported by the ARO Grants No. W911NF-21-2-0162 (ETHOS) and W911NF-21-1-0113, and the Vannevar Bush Faculty Fellowship (VBFF) grant no. N00014-20-1-2834 from the Department of Defense.  C.P.  acknowledges the support from a public grant overseen by the French Agence Nationale de la Recherche under grant agreement no. ANR-21-CE24-0032 (SUPERSPIN). We also acknowledge the computational support from the Arkansas High Performance Computing Center (AHPCC) for computational resources.	
\end{acknowledgements}

	\bibliography{PRB}% Produces the bibliography via BibTeX.

\pagebreak
\widetext
\begin{center}
	\textbf{\large Supplemental Materials for ``Tuning photostriction in (PbTiO\ensuremath{_{3}})\ensuremath{_{n}}/(SrTiO\ensuremath{_{3}})\ensuremath{_{m}} superlattices via chemical composition: An \textit{ab-initio} study"}
\end{center}

\section{Average c/a in the $(1|9)$, $(4|6)$, $(6|4)$ and $(9|1)$ SLs}
\begin{figure}[ht!]
	\centering
	\includegraphics[width=.85\linewidth]{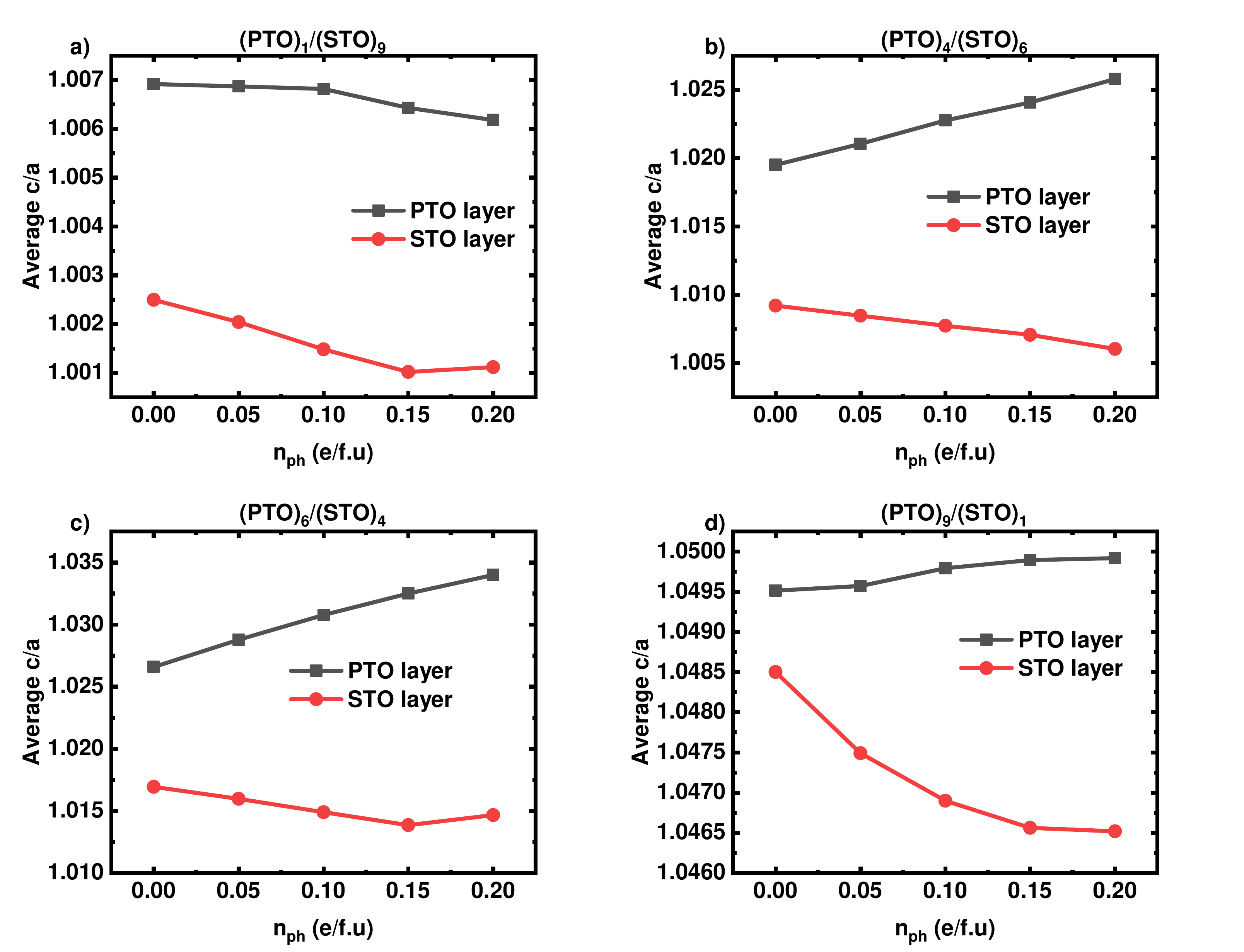}
	\caption{a)  Average c/a for PTO and STO layer in the $(1|9)$ SLs, b)  Average c/a for PTO and STO layer in the $(4|6)$ SLs, c)  Average c/a for PTO and STO layer in the $(6|4)$ SLs, d)  Average c/a for PTO and STO layer in the $(9|1)$ SLs  as function of $n_{ph} = 0.0-0.2$ e/f.u.}
\end{figure}

\newpage
\section{Polarization in the $(1|9)$, $(4|6)$, $(6|4)$ and $(9|1)$ SLs}
\begin{figure}[ht!]
	\centering
	\includegraphics[width=.85\linewidth]{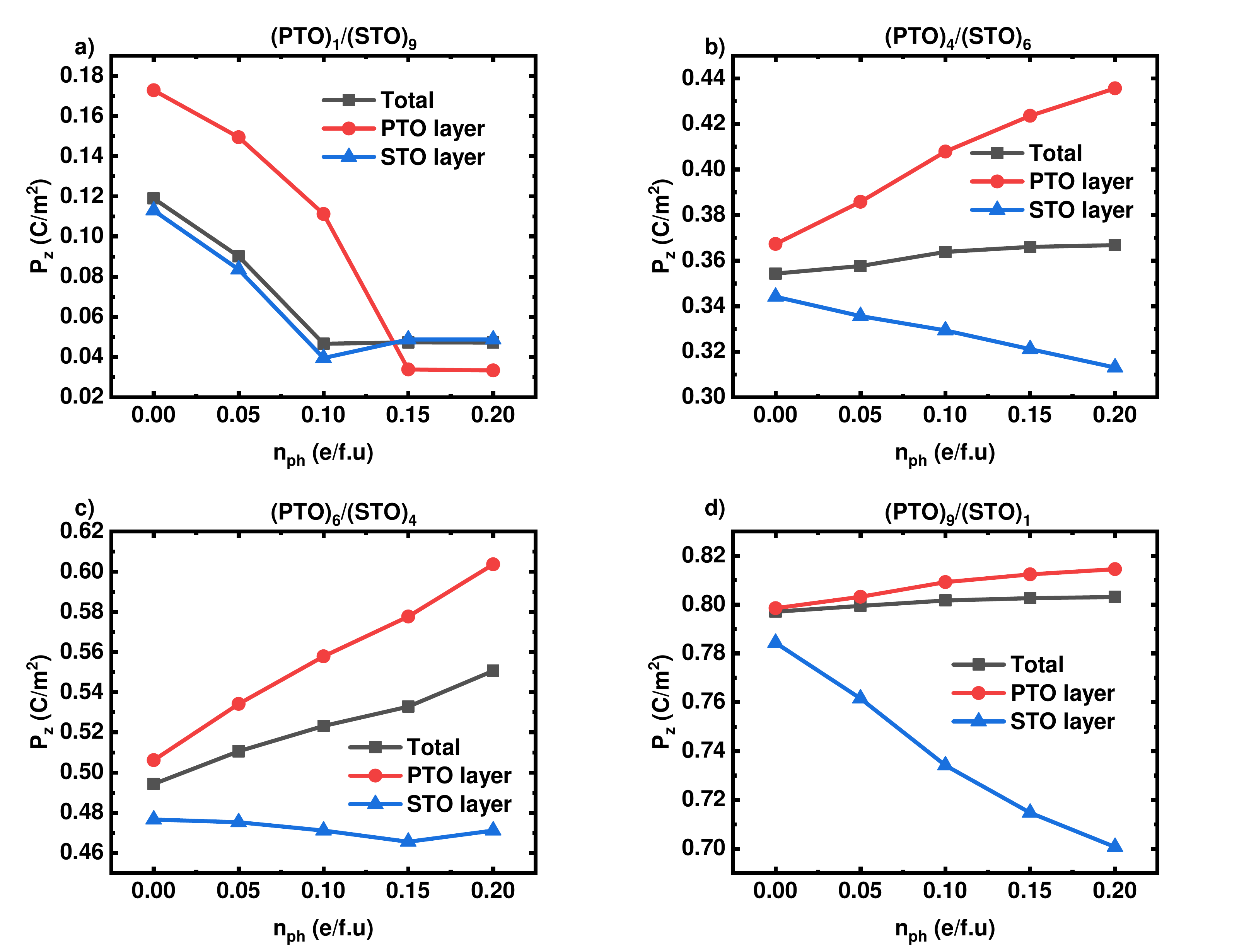}
	\caption{Polarizations as function of number of excited carriers for the a) $(1|9)$ SLs, b) $(4|6)$ SLs, c) $(6|4)$ SLs, d) $(9|1)$ SLs.}
\end{figure}
\newpage

\section{Values of $\alpha$ for each of the considered SLs}
\begin{figure}[ht!]
	\centering
	\includegraphics[width=0.5\linewidth]{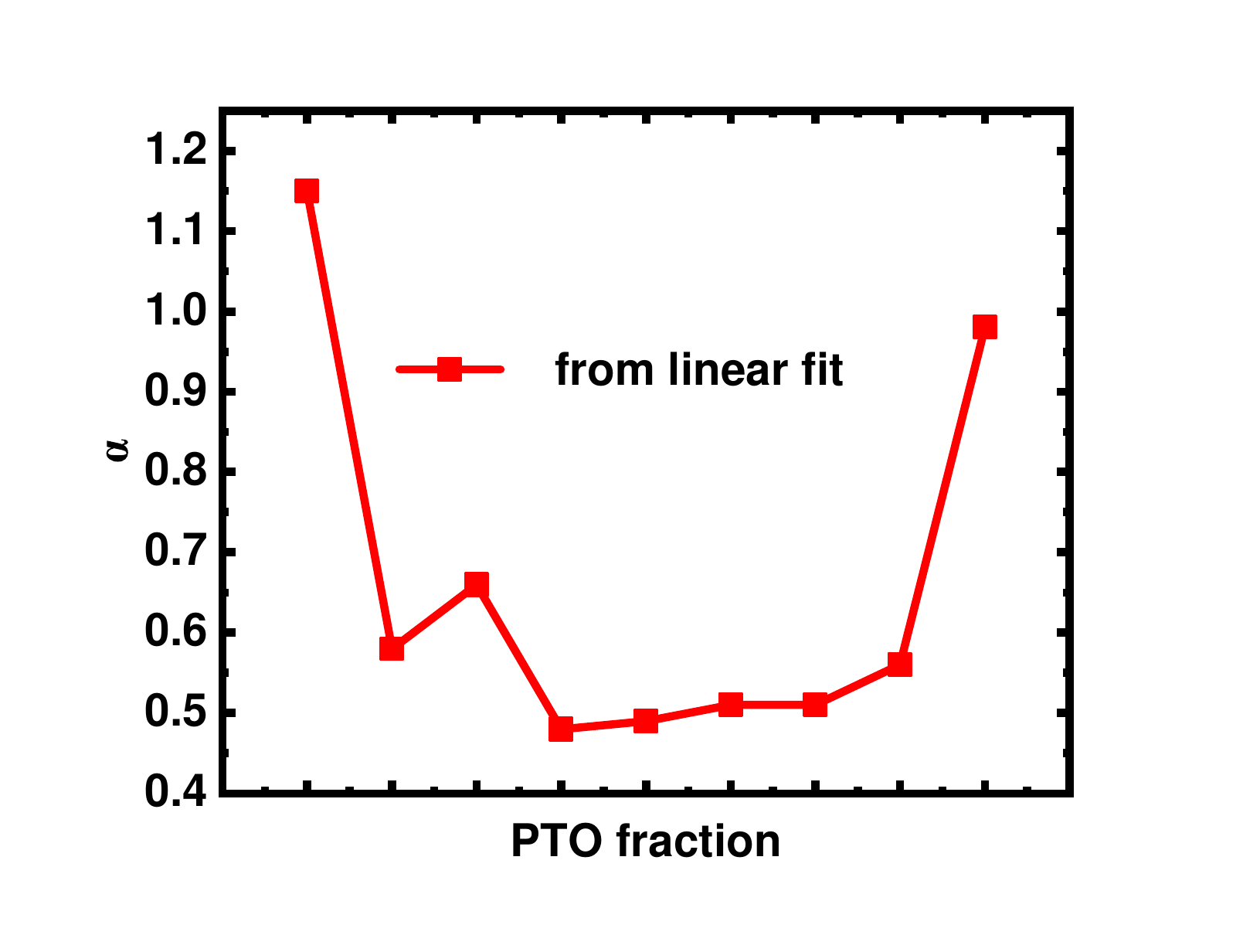}
	\caption{a) Values of $\alpha$ as function of the $(n|n)$ SLs; b)  Values of $\alpha$ as function of PTO fraction in the $(n|m)$ SLs.}
\end{figure}

\section{results of the model using $P_{z} \propto \eta_{T}^{\alpha} $ for the $(1|9)$, $(4|6)$, $(6|4)$ and $(9|1)$ SLs}
\begin{figure}[ht!]
	\centering
	\includegraphics[width=.75\linewidth]{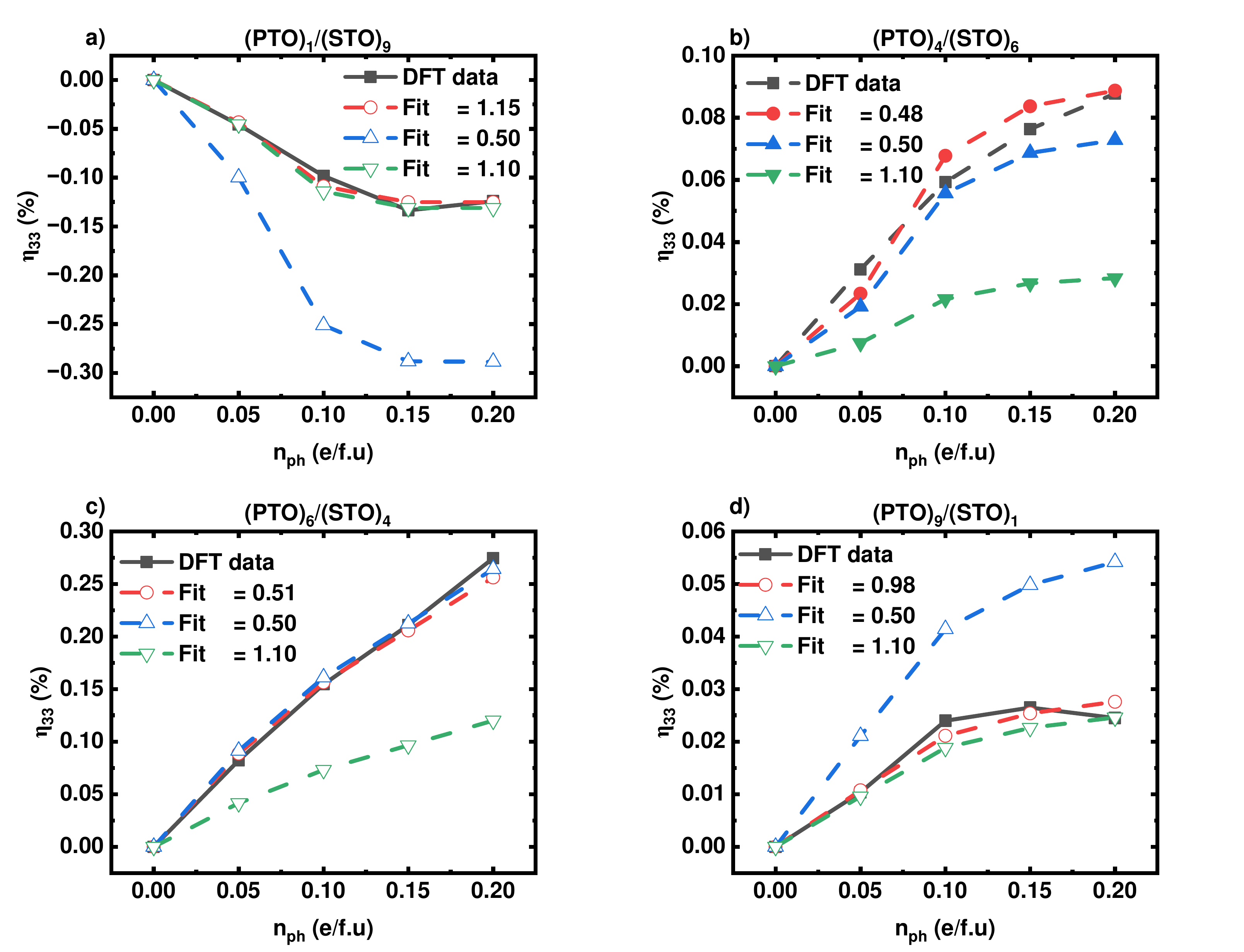}
	\caption{Different fits of the DFT data ($\eta_{33}~ vs~ n_{ph}$) using the ``$P_{z} \propto \eta_{T}^{\alpha} $ model" in the main text for a) $(1|9)$ SLs, b) $(4|6)$ SLs, c) $(6|4)$ SLs, d) $(9|1)$ SLs.}
\end{figure}

\newpage

\section{Results of the converse piezoelectric for the $(1|9)$, $(4|6)$, $(6|4)$ and $(9|1)$ SLs}

\begin{figure}[ht!]
	\centering
	\includegraphics[width=.85\linewidth]{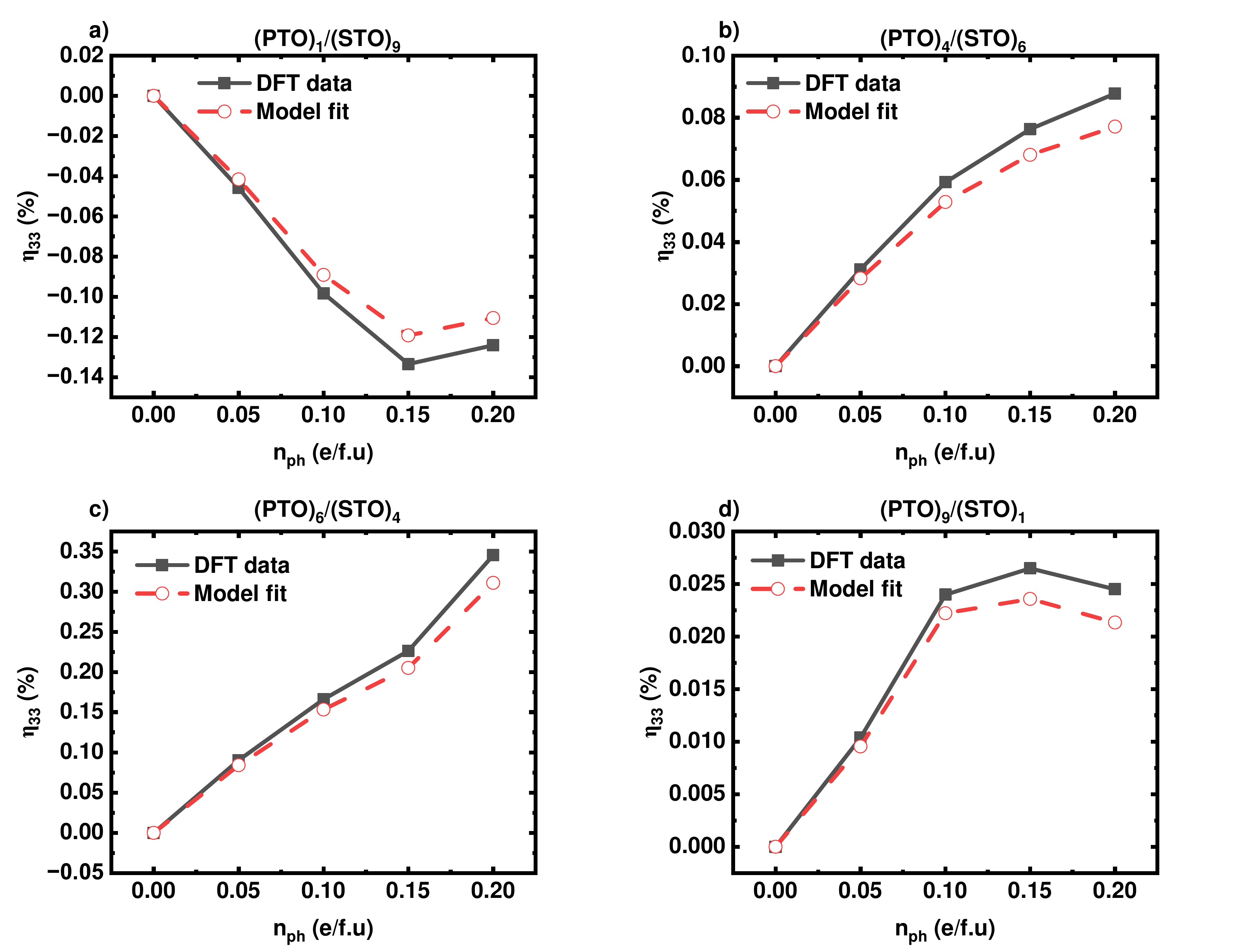}
	\caption{Different fits of the DFT data ($\eta_{33}~ vs~ n_{ph}$) using the converse piezoelectric model in the main text for a) $(1|9)$ SLs, b) $(4|6)$ SLs, c) $(6|4)$ SLs, d) $(9|1)$ SLs.}
\end{figure}
\newpage
\section{Expression of the total strain in the SLs}
\begin{figure*}[ht!]
	\centering
	\includegraphics[width=1\linewidth]{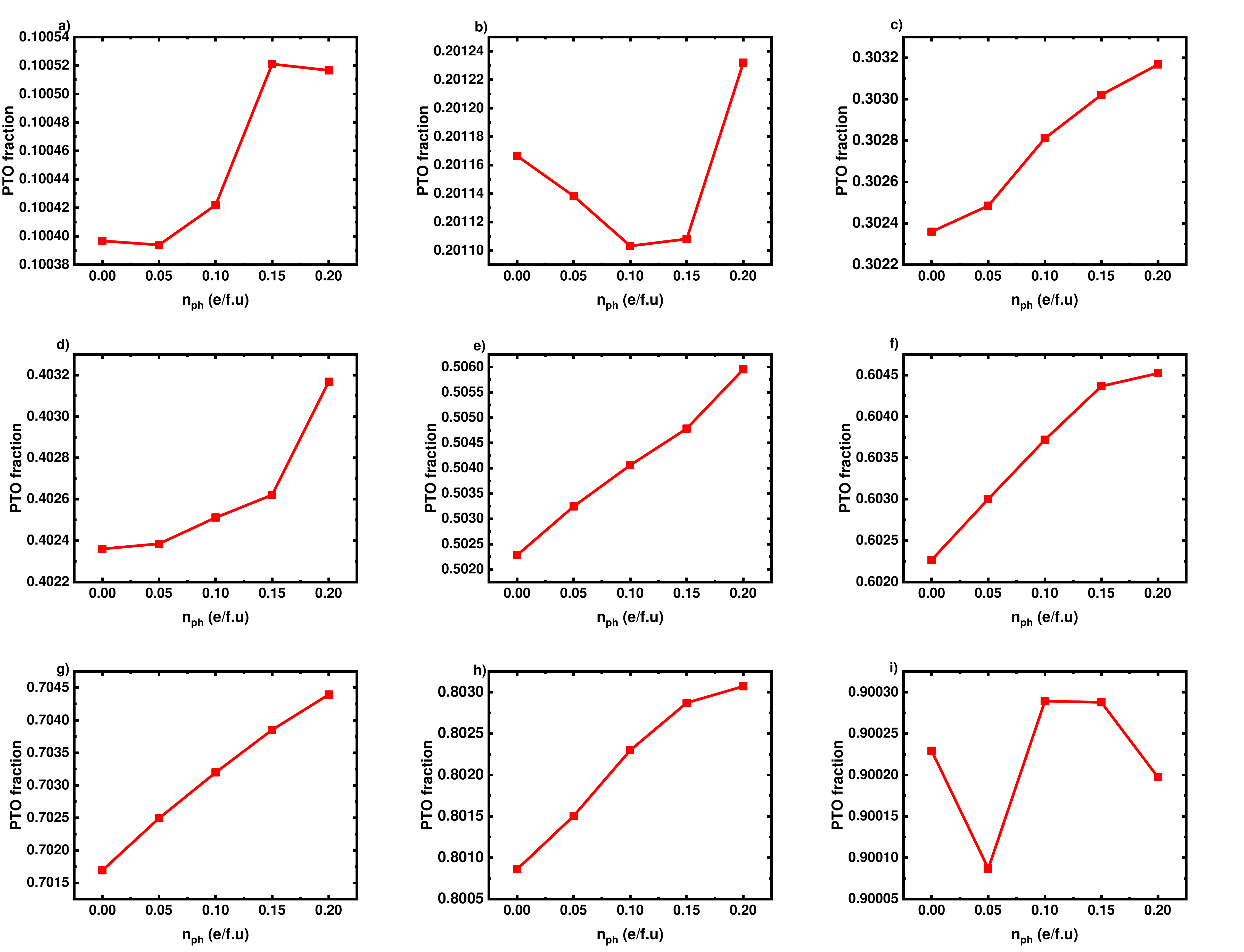}
	\caption{\label{fa} Variation of the volume fraction $x$ as function of $n_{ph}$ for each of the SLs. a) to i) correspond to the $(1|9)$, $(2|8)$,$(3|7)$,$(4|6)$,$(5|5)$,$(6|4)$,$(7|3)$,$(8|2)$,$(9|1)$ respectively.}
\end{figure*}

Let $l_{p}$ and $l_{s}$	be the lengths of the PTO and STO layer in the ground state structures ($n_{ph} = 0$). Also let  $l_{p}^{l}$ and $l_{s}^{l}$ be their lengths after illumination ($n_{ph} \neq 0$). The light-induced strain in the SLs is as follows:
\begin{equation}
	\eta_{33}= \dfrac{l_{s}^{l}+l_{p}^{l} -l_{s}-l_{p}}{l_{s}+l_{p}} = \dfrac{l_{s}^{l} -l_{s}}{l_{s}+l_{p}} + \dfrac{l_{p}^{l}-l_{p}}{l_{s}+l_{p}};
\end{equation}
now we write:
\begin{equation}
	\dfrac{l_{s}^{l} -l_{s}}{l_{s}+l_{p}} = \dfrac{l_{s}}{l_{s}+l_{p}} 	\dfrac{l_{s}^{l} -l_{s}}{l_{s}};~~ \dfrac{l_{p}^{l}-l_{p}}{l_{s}+l_{p}} =  \dfrac{l_{p}}{l_{s}+l_{p}} \dfrac{l_{p}^{l}-l_{p}}{l_{p}}.
\end{equation}
In the optimized ground state structure, the volume fraction $x$ of PTO and $(1-x)$ of STO in the SLs can be calculated by the followings:
\begin{equation}
	x = \dfrac{l_{p}}{l_{s}+l_{p}}  \mbox{ and } 1- x = \dfrac{l_{s}}{l_{s}+l_{p}}.
\end{equation}
Defining the following induced-strains in PTO and STO layers:
\begin{equation}
	\eta_{33}^{p} = \dfrac{l_{p}^{l}-l_{p}}{l_{p}} \mbox{ and } \eta_{33}^{s} = \dfrac{l_{s}^{l}-l_{s}}{l_{s}},
\end{equation}
we arrive at the formula for Eq. 2 in the main text:
\begin{equation}\label{a1}
	\eta_{33} \approx x\eta_{33}^{p} + (1-x)\eta_{33}^{s}
\end{equation}
Note that in this derivation, $x$ as defined in Eq. 3 is independent of $n_{ph}$. However, the lengths of the PTO and STO layer of the superlattices obtained from DFT change with $n_{ph}$ and thus the variation of $x$ as function of $n_{nph}$ as shown on Figs \ref{fa}.  That is why we carefully use the approximation sign in written the final expression in Eq. \ref{a1}. To have the converse piezoelectric model to better agree with the DFT data, as presented in this paper, we use the values of $x$ as shown on Figs \ref{fa}. 
\section{Details on the estimation of the $r^{i}$ ratio from the DFT data}
Following \cite{sri202}, for each SLs, the polarization in each layer can be related to its tetragonal distortion via the following relation:
\begin{equation}\label{b1}
	P_{i}^{2} \propto \dfrac{c_{i}}{a}-1
\end{equation}
with $c_{i} $ being the average out-of-plane lattice parameter of the layer of the SLs ($i=$1 for PTO and 2 for STO). Note that $c_{i}$ is function of $n_{ph}$. Taking the logarithmic derivative of Eq.(\ref{b1}) gives:
\begin{equation}\label{b2}
	2 \dfrac{\delta P_{i}}{P_{i}} = \dfrac{\delta c_{i}}{c_{i}-a},
\end{equation} 
which can be rewritten as:
\begin{equation}\label{b3}
	\delta P_{i} = \dfrac{c_{i} P_{i}}{2} \dfrac{\eta_{i}}{c_{i}-a},
\end{equation}
writting $\delta P_{i} = \epsilon_{0} \chi_{33}^{i}\delta E_{i}$ and $\eta_{i} = d_{33}^{i}\delta E_{i}$, we have:
\begin{equation}
	\epsilon_{0} \chi_{33}^{i} = \dfrac{c_{i} P_{i}}{2} \dfrac{d_{33}^{i}}{c_{i}-a}
\end{equation}
and thus we get:
\begin{equation}
	r^{i}= \dfrac{d_{33}^{i}}{\chi_{33}^{i}} = 2 \epsilon_{0} \dfrac{c_{i}-a}{c_{i} P_{i}}
\end{equation}

\end{document}